\documentstyle[epsfig,12pt]{article}
\newcommand{\BE}{\begin{equation}}
\newcommand{\EE}{\end{equation}}
\begin{document}
\begin{center}
{\Large{\bf Effect of asynchronicity on the universal behaviour of
coupled map lattices}}\\
  \vspace{1cm} Neelima Gupte$^{a,1}$, T.M. Janaki$^{b,2}$ and Sudeshna
  Sinha$^{b,3}$\\ $^a$ {\it Department of Physics, IIT Madras, Chennai
    600036, India}\footnote{gupte@chaos.iitm.ernet.in}\\ $^b${\it
      Institute of Mathematical Sciences, Taramani, Chennai 600 113,
      India}\footnote{janaki@imsc.ernet.in}$^, $\footnote{sudeshna@imsc.ernet.in}
\end{center}

\begin{abstract}
  
  We investigate the spatiotemporal dynamics of coupled circle map
  lattices, evolving under synchronous (parallel) updating on one hand
  and asynchronous (random) updating rules on the other.  Synchronous
  evolution of extended spatiotemporal systems, such as coupled circle
  map lattices, commonly yields multiple co-existing attractors,
  giving rise to phenomena strongly dependent on the initial lattice.
  By marked contrast numerical evidence here strongly indicates that
  asynchronous evolution eliminates most of the attractor states
  arising from special sets of initial conditions in synchronous
  systems, and tends to yield more global attractors. Thus the
  phenomenology arising from asynchronous evolution is more generic
  and robust in that it is obtained from many different classes of
  initial states. Further we show that in parameter regions where both
  asynchronous and synchronous evolution yield spatio-temporal
  intermittency, asynchronicity leads to better scaling behaviour.
\end{abstract}

\newpage

It is well-known that spatially extended systems undergoing temporal
evolution show the presence of a large number of multiple co-existing
attractors \cite{attract}. In the case of coupled map lattices, which
constitute simple models of spatially extended systems
\cite{cml,kaneko} evolving under synchronous up-dates, it has been
seen in numerous examples that system attractors shows strong
sensitivity to initial conditions even in parameter regimes where the
evolving dynamics is spatio-temporally periodic in nature
\cite{gauri}. This multiattractor property has significant
consequences for problems like control and synchronisation \cite{us}
in physical, chemical, biological and engineering contexts. The
stability of such attractors to perturbations and noise have also been
studied in the case of globally coupled systems \cite{kaneko}.

A typical CML consists of dynamical elements on a lattice which
interact with suitably chosen sets of other elements, and evolve via
synchronous or parallel updates wherein the dynamical elements at each
lattice site are updated simultaneously.  However, there have been
several attempts to study CML-s which evolve via asynchronous
evolution, that is, one in which the updates at lattice sites are not
concurrent, but sequential instead. The study of asynchronous updates
is considered to be interesting for several reasons. Notably,
neurophysiological systems like neurons and neuron groups evolve
asynchronously, and lattice dynamical models of such phenomena must of
course employ asynchronous updating schemes. Therefore it is of
importance to investigate the effects of asynchronicity in prototype
models [6-13]. These effects can be quite significant, e.g.  it has
been argued that asynchronous up-dates can alter the universality
class of spatio-temporal intermittency \cite{asyn3}.

In this paper, we observe that asynchronicity can have a very
important physical effect. Systems which show the existence of
multiple co-existing attractors which depend on the state of the
initial lattice when evolved via synchronous updates, evolve to the
same global attractor from many kinds of initial states under
asynchronous evolution. Thus asynchronicity wipes out multiple
co-existing attractors and leads to a generic global attractor which
is robust to the evolution of different classes of initial conditions.

We demonstrate this result for the following specific example.  The
space on which our CML dynamics occurs is a discrete $1$-dimensional
chain, with periodic boundary conditions. The sites are denoted by
integers $i$, $i = 1, \dots, N$, where $N$ is the linear size of the
lattice. On each site is defined a continuous state variable denoted
by $x_n (i)$, which corresponds to the physical variable of interest,
with $n$ denoting the discrete time. Here the local on-site map is
chosen to be the sine circle map, a map which has generated much
research interest:
$$f(x) = x + \omega - \frac{K}{2 \pi} \sin (2 \pi x)$$ where $ 0 \leq
x \leq 1$.  The parameter $K$ indicates the strength of the
nonlinearity, and is chosen to be $1$ here.

These local maps are coupled through their nearest neighbours, and the
coupling form is the discretized Laplacian form (i.e. future coupled)
\cite{cml}, with the strength of coupling given by $\epsilon$. Now
extensive results have been obtained for the standard parallel time
evolution of lattices of circle maps, i.e. the scheme in which all
individual maps of the lattice are iterated forward simultaneously
\cite{gauri}. This synchronous evolution implies the dynamics: \BE
x_{n+1} (i) = (1 - \epsilon) f(x_n (i)) + \frac{\epsilon}{2}\{ f(x_n
(i-1)) + f(x_n (i+1)) \}  \ \ \ \ \rm{mod}
\ 1\EE

It is known \cite{cml, gauri} that varying $\epsilon$ and $\omega$,
under the usual parallel undating, yields various dynamical phases,
such as: the synchronized fixed point, i.e. spatial period 1 temporal
period 1 (S1T1), spatio-temporally periodic solutions e.g. spatial
period 2 temporal period 1 (S2T1) and spatial period 2 temporal period
2 (S2T2), spatial intermittency, spatiotemporal intermittency etc. For
parallel updating an important feature of the spatiotemporal dynamics
is the existence of multiple co-existing attractors. This implies that
the resultant attractor is very strongly dependent on the
spatiotemporal features of the initial lattice. Thus synchronous
evolution can yield many different behaviours under different initial
preparations even for identical parameter values.  Detailed phase
diagrams of the behaviour of this CML under synchronous evolution have
been obtained\cite{gauri,gauri1}.

Here we focus on randomly updated CMLs, i.e. a CML where the elements
of the lattice do not update simultaneously, but {\em update one after
  another in random sequence}. Time now is measured in units of one
complete updating sweep of the lattice (in random order), i.e.  after
one unit of time, all sites in the lattice have been updated. This
random updating scheme should help us test the implications of
asynchronicity in extended systems, and help us gauge the degree of
robustness of various physical features emerging under conventional
synchronous evolution.

We study the different $\epsilon$ and $\omega$ regions of this coupled
circle map system, under three classes of initial lattices: (i)
spatial period 2 initial condition; (ii) spatial period 2 initial
lattice, with a kink; and (iii) random initial lattice. We observe the
spatiotemporal dynamics under parallel updating on one hand and
completely asynchronous updating on the other, and find out how this
``phase diagram'' changes under asynchronicity, for the
different classes of initial states mentioned above.\\

{\em Case I:} In the region with very high coupling parameter
$\epsilon$ and low $\omega$, for instance around $\epsilon \sim 0.97$,
$\omega \sim 0.02$, parallel updating yields spatial period 2,
temporal period 2 behaviour when evolved from spatial period 2 initial
lattices. But for the case of the initial periodic lattice having a
kink, and for the case of random initial lattices, spatio-temporal
intermittency is obtained. (See \cite{gauri} for a detailed phase
diagram).

By contrast, under asynchronous updating in this parameter regime one
obtains spatiotemporal fixed points for {\em all} initial conditions:
spatial period 2, spatial period 2 with a kink, and random initial
lattices. This fixed point is at $\theta = f(\theta) \sim \omega$ for
low $\omega$. (See Fig.~1.) Note that this is in agreement with
earlier observations that under strong coupling asynchronicity has the
effect of regularising the system \cite{asyn8}.

Thus, under randomness in updating rules, spatiotemporal dynamics is
no longer sensitive to initial conditions. The spatiotemporal fixed
point is now the global attractor of the system.\\

{\em Case II:} In the region around $\epsilon \sim 0.8$, $\omega \sim
0.2$, asynchronicity yields spatiotemporal chaos from all initial
conditions. By contrast, parallel updating yields spatiotemporal
periodicity when evolved from spatial period 2 initial lattices (with
or without kink), and yields spatiotemporal intermittency when evolved
from random initial lattices (see Fig.~2).

Also note here, that while asynchronicity had the effect of
regularising the system under strong coupling (as seen above in Case
I), under weaker coupling asynchronicity has the effect of inducing
spatiotemporal disorder. In this example for instance, we see that
parallel updating gives rise to spatiotemporal periodicity for certain
classes of initial states, while asynchronous evolution always
leads to spatiotemporal chaos.\\

{\em Case III:} Around the region $\epsilon \sim 0.8$, $\omega \sim
0.08$, asynchronicity yields spatiotemporal intermittency for all
initial conditions. In contrast, parallel updates yield exact
spatiotemporal periodicity for period 2 initial lattices, and
spatiotemporal intermittency for random lattices.

Interestingly, when both asynchronous and synchronous evolution yield
spatiotemporal intermittency from random initial lattices, the scaling
exponents obtained from the probability distribution of laminar
lengths in space and time are quite distinct for the two cases. The
intermittent dynamics arising from asynchronous evolution exhibits
scaling in both time and space, with the spatial and temporal scaling
exponents being approximately the same ($\phi \sim 3$). Synchronous
evolution on the other hand leads only to temporal scaling, and not
good spatial scaling (see Fig.~3).\\

We note that similar phenomena are observed for coupled logistic map
lattices as well, i.e. a CML with the local on-site map being $f(x) =
r x(1-x)$, with $r=4$. See figures 6 and 7 for two examples, one for
high coupling strength $\epsilon = 0.9$ and the other for low coupling
strength $\epsilon =0.1$. Clearly here too asynchronous evolution is
insensitive to differences in initial conditions, whereas multiple
attractors co-exist for the case of the traditional synchronous
evolution.

In summary then, we have investigated the spatiotemporal dynamics of
coupled map lattices evolving under asynchronous updating rules. We
have shown that asynchronicity has a pronounced effect on the
spatiotemporal dynamics: it yields a more global attractor where
multiple attractors co-existed for the case of parallel updating. In
that sense asynchronous updating yields more generic and robust
phenomena in extended systems. Conversely, introducing some degree of
asynchronicity in an extended system may help lead the system to its
most generic attractor from any initial condition, however special.
The nature of the generic attractor i.e. whether regular or
disordered, appears to depend on the strength of the coupling, as well
as the synchronicity /asynchronicity. In the case of spatio-temporal
intermittency, asynchronicity leads to better spatial scaling. Thus,
synchronous and asynchronous updates can lead to different
universality classes of spatio-temporal behaviour for systems which
are otherwise identical in all respects.  Therefore asynchronicity
constitutes a relevant perturbation in the evolution of extended
systems, and can perhaps be used to direct spatially extended systems
to desired global attractors.  We hope our observations will be useful
towards the understanding the diversity of phenomena which can be seen
in spatially extended systems and for controlling their behaviour.

\newpage
\pagestyle{empty}
\begin{figure}
\mbox{\epsfig{file=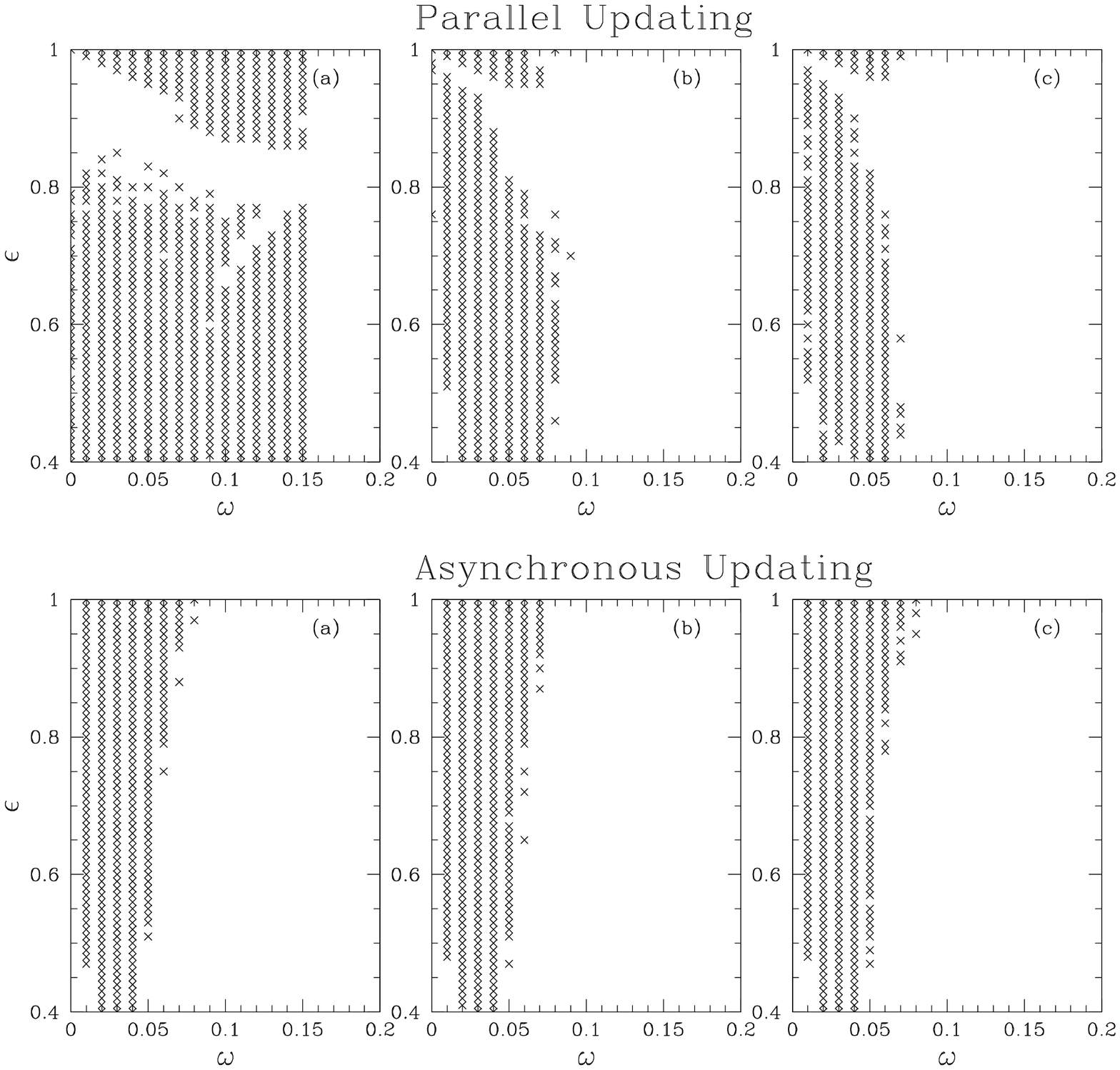,width=12truecm}}
\caption{Area of $\epsilon$-$\omega$ parameter space occupied by the
  spatiotemporal fixed point for the case of parallel updating (top
  panel) and completely asynchronous updating (bottom panel), for
  different initial conditions: (a) period 2 initial lattice (b)
  period 2 with a kink and (c) random initial conditions.  Here
  lattice size $N = 100$.}
\end{figure}

\begin{figure}
\mbox{\epsfig{file=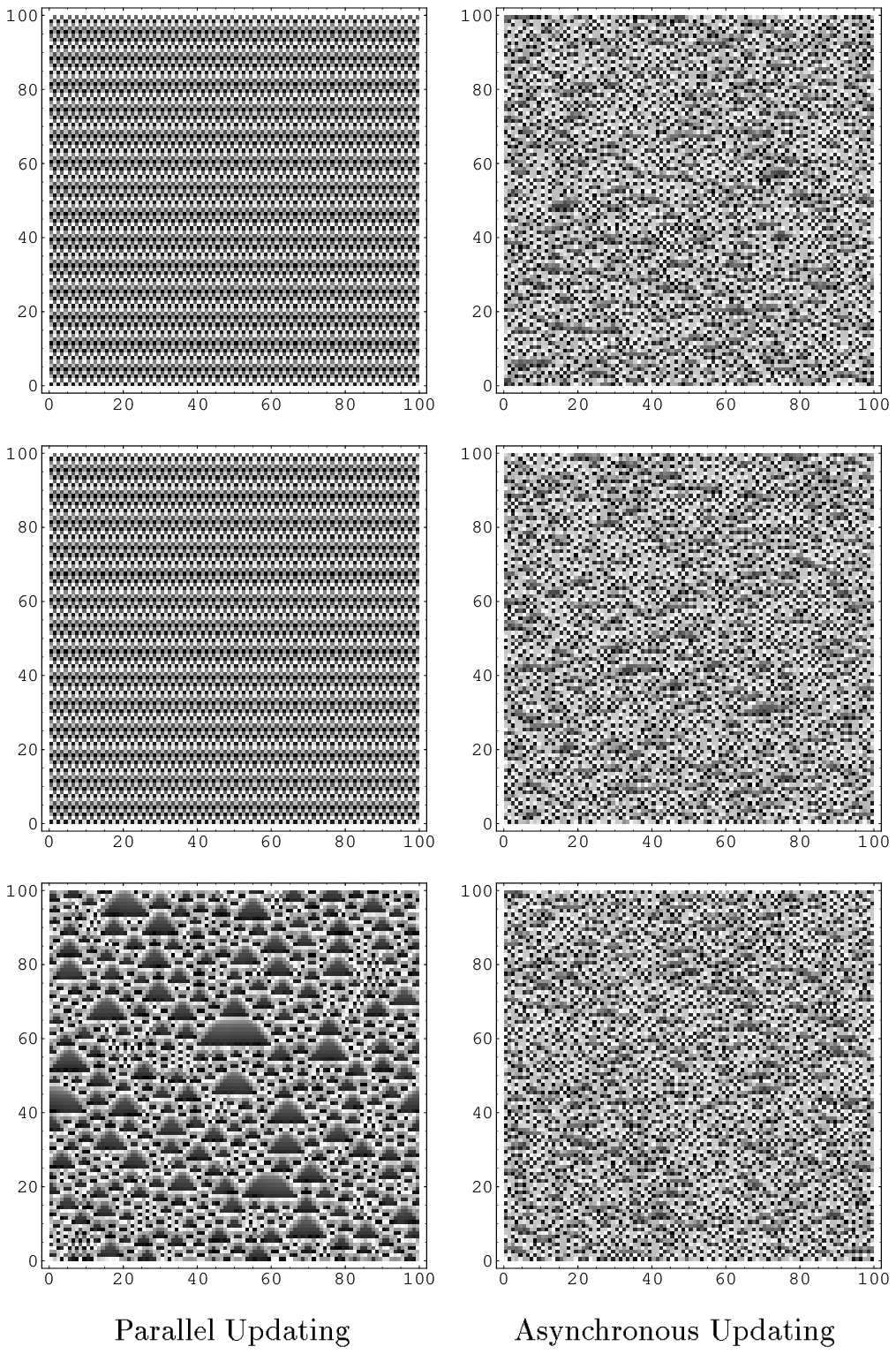,width=19truecm}}
\end{figure}

\begin{figure}
\caption{Space-time density plots showing the evolution of a circle
  map lattice of size $N = 100$ over 100 iterations via parallel
  updating and via completely asynchronous updating. Each slice
  parallel to the $x$ axis displays a snap shot of the spatial profile
  at some instant of time. Here $\epsilon = 0.8, \omega = 0.2$. The
  lattices are evolved from (top to bottom) period 2 initial lattice,
  period 2 lattice with a kink at $i = 37$ and random initial
  lattice.}
\end{figure}

\begin{figure}
\mbox{\epsfig{file=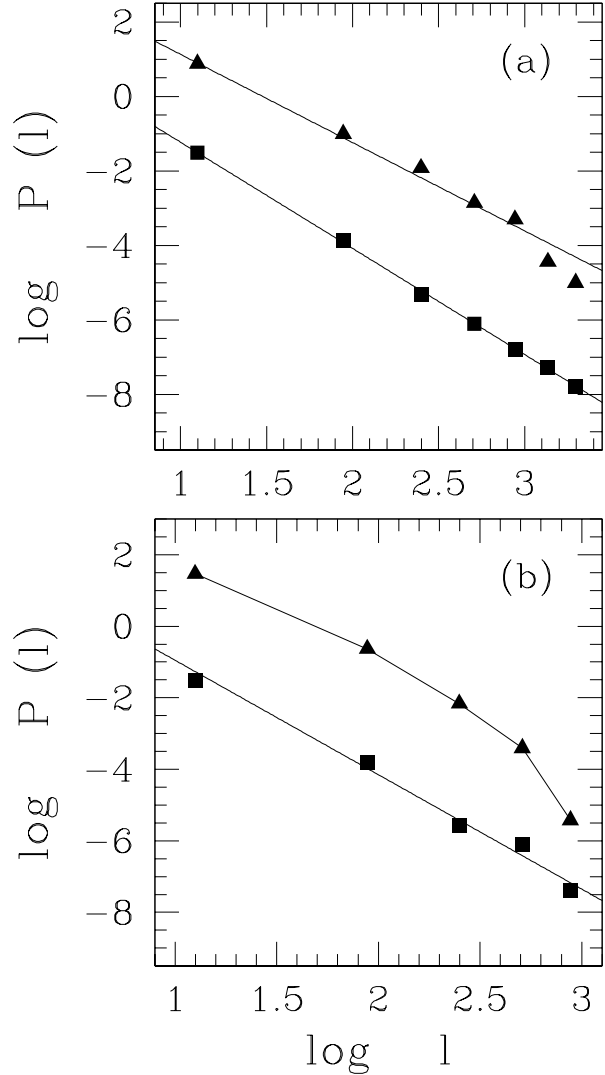,width=15truecm}}
\caption{
  Probability $P$ of occurence of (a) temporal laminar regions and (b)
  spatial laminar regions vs length of laminar regions $l$ (on a
  log-log plot), for the case of asynchronous updating (solid squares)
  and parallel updating (solid triangles). The solid lines are the
  best fit straight lines, with slope equal to $-3.0$ for asynchronous
  updating and $-2.4$ for parallel updating in (a), and with slope
  $-3.2$ for asynchronous updating in (b). These results are for
  circle map lattices of size $N = 1000$, with $\epsilon = 0.8, \omega
  = 0.08$, evolved from random initial conditions. Sample size is
  $\sim 10^4$.}
\end{figure}

\begin{figure}
\mbox{\epsfig{file=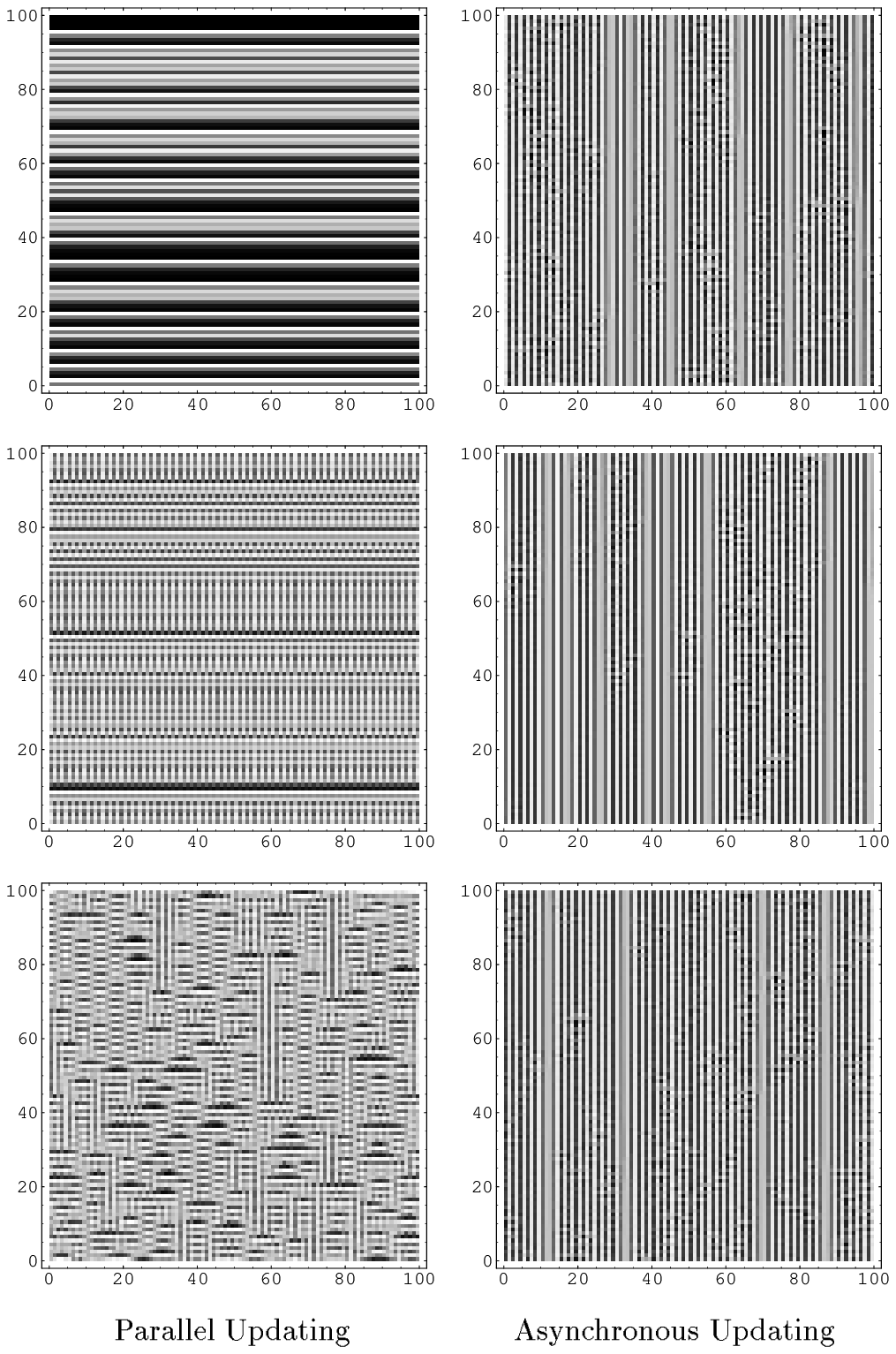,width=19truecm}}
\end{figure}

\begin{figure}
\caption{Space-time density plots showing the evolution of a logistic
  map lattice of size $N = 100$ over 100 iterations via parallel
  updating and via completely asynchronous updating. Each slice
  parallel to the $x$ axis displays a snap shot of the spatial profile
  at some instant of time. Here $\epsilon = 0.9$. The lattices are
  evolved from (top to bottom) uniform (spatial period 1) initial
  lattice, spatial period 2 initial lattice and random initial
  lattice.}
\end{figure}

\begin{figure}
\mbox{\epsfig{file=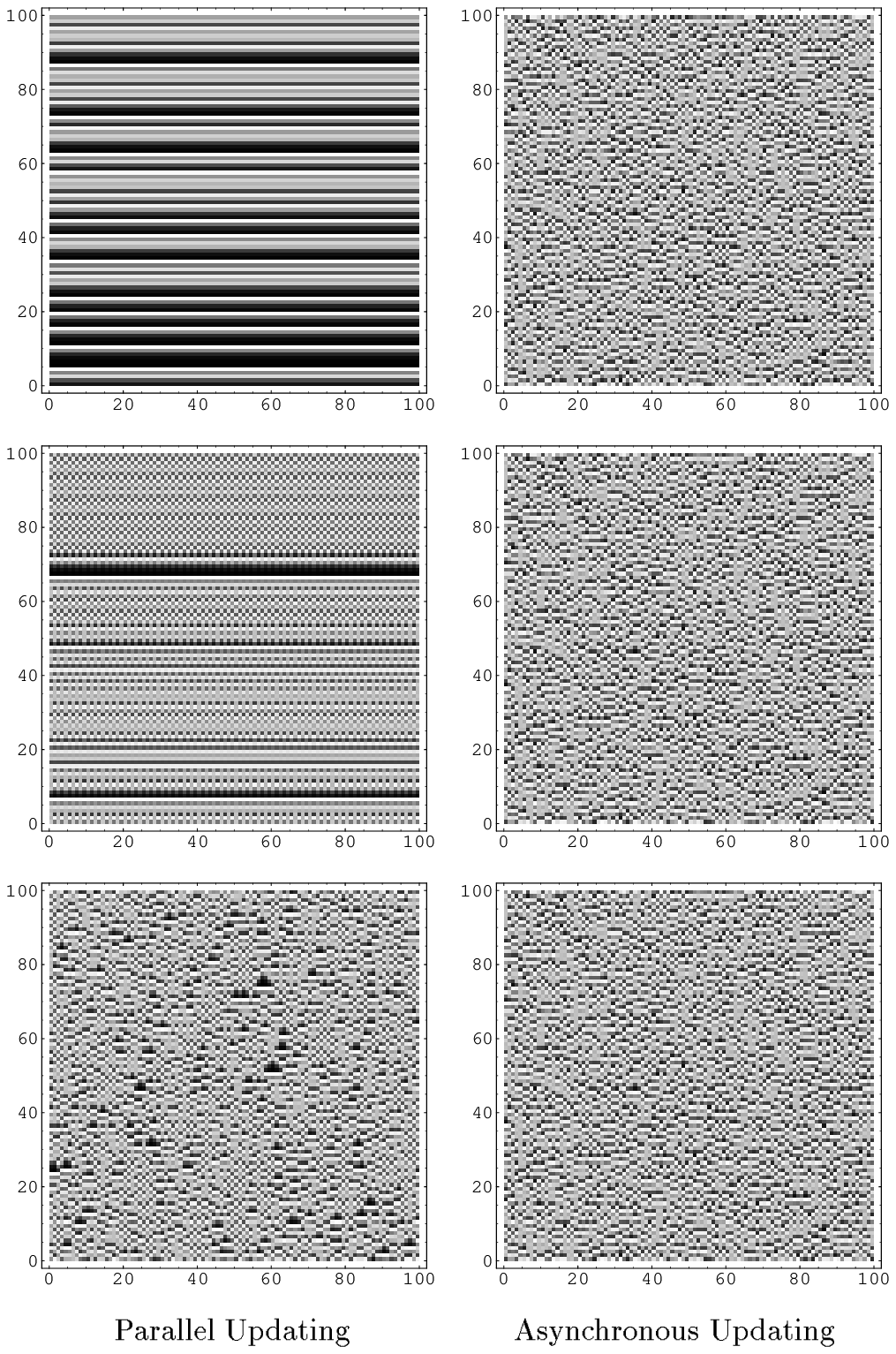,width=19truecm}}
\end{figure}

\begin{figure}
\caption{Space-time density plots showing the evolution of a logistic
  map lattice of size $N = 100$ over 100 iterations via parallel
  updating and via completely asynchronous updating. Each slice
  parallel to the $x$ axis displays a snap shot of the spatial profile
  at some instant of time. Here $\epsilon = 0.1$. The lattices are
  evolved from (top to bottom) uniform (spatial period 1) initial
  lattice, spatial period 2 initial lattice and random initial
  lattice.}
\end{figure}

\end{document}